\documentclass[a4paper]{elsarticle}
\usepackage{graphicx}
\usepackage{subfigure}
\usepackage{setspace}
\usepackage{textcomp}
\usepackage{SIunits}
\usepackage{layout}
\usepackage{textcomp}
\usepackage{tabularx,multirow,booktabs,blindtext}
\usepackage{multirow}
\usepackage{kantlipsum}
\usepackage{multicol,caption}
\newenvironment{Figure}
  {\par\noindent\minipage{\linewidth}}
  {\endminipage\par}

\begin{document}
\large

\title{Interaction between U/UO$_{2}$ bilayers and hydrogen studied by in-situ X-ray diffraction}

\author{J E Darnbrough$^{1}$, R. M. Harker$^{2}$, I Griffiths$^{1}$, D. Wermeille$^{3}$, G. H. Lander$^{4}$ and R. Springell$^{1}$}

\address{$^1$University of Bristol, Interface Analysis Centre, HH Wills Physics Laboratory, Tyndall Avenue, Bristol, BS2 8BS UK}

%\address{$^2$University of Oxford, Department of Materials, Parks Road, Oxford, OX1 3PH, UK}

\address{$^2$AWE Aldermaston, Metal Chemistry, Reading, RG74PR UK} 

\address{$^3$XMaS, European Synchrotron Radiation Facility, BP220, F-38043 Grenoble, France} 

\address{$^4$European Commission, Joint Research Centre, Directorate for Nuclear Safety and Security, Postfach 2340, D-76125 Karlsruhe, Germany}

\ead{ed.darnbrough@materials.ox.ac.uk}

\begin{abstract}
This paper reports experiments investigating the reaction of H$_{2}$ with uranium metal-oxide bilayers. The bilayers consist of $\leq$ 100 nm of epitaxial $\alpha$-U (grown on a Nb buffer deposited on sapphire) with a UO$_{2}$ overlayer of thicknesses of between 20 and 80 nm. The oxides were made either by depositing via reactive magnetron sputtering, or allowing the uranium metal to oxidise in air at room temperature. The bilayers were exposed to hydrogen, with sample temperatures between 80 and 200 C, and monitored via in-situ x-ray diffraction and complimentary experiments conducted using Scanning Transmission Electron Microscopy - Electron Energy Loss Spectroscopy (STEM-EELS). Small partial pressures of H$_{2}$ caused rapid consumption of the U metal and lead to changes in the intensity and position of the diffraction peaks from both the UO$_{2}$ overlayers and the U metal. There is an orientational dependence in the rate of U consumption. From changes in the lattice parameter we deduce that hydrogen enters both the oxide and metal layers, contracting the oxide and expanding the metal. The air-grown oxide overlayers appear to hinder the H$_{2}$-reaction up to a threshold dose, but then on heating from 80 to 140 C the consumption is more rapid than for the as-deposited overlayers. STEM-EELS establishes that the U-hydride layer lies at the oxide-metal interface, and that the initial formation is at defects or grain boundaries, and involves the formation of amorphous and/or nanocrystalline UH$_{3}$. This explains why no diffraction peaks from UH$_{3}$ are observed. \footnotetext[2]{\textcopyright British Crown Owned Copyright 2017/AWE}
\end{abstract}

\maketitle

\begin{multicols}{2}

\section{Introduction}
The safe, long term, storage of metallic uranium samples requires a fundamental understanding and predictive capability regarding the uranium-hydrogen reaction \cite{Glascott2013}. Its storage in the presence of either moisture or organic material in sealed containers is known to produce hydrogen over time, which goes on to react with the uranium.  As described elsewhere, the reaction proceeds through four characteristic phases: induction, acceleration, linear and terminal \cite{Harker2006} and the last three phases can produce finely divided, highly reactive (pyrophoric) radioactive powder in the form of UH$_{3}$. Generally a mixture of $\alpha$ and $\beta$-UH$_{3}$ is formed between ambient temperatures and 100 C with the concentration of $\alpha$-UH$_{3}$ increasing to significant proportions at room temperature and below \cite{Orr2016}. This uranium hydride presents a significant safety hazard on opening the container to air and knowing the amount of hydride produced allows a quantification of the hazard \cite{Glascott2003}.

There are, however, many unanswered questions remaining, despite a fairly well-developed literature, and at least part of the problem stems from the complexity of the real-world metallic uranium system (containing inclusions, defects, stress and many oxide variables etc). In this work, we have attempted to simplify the system as much as possible. To this end we have prepared model UO$_{2}$/U metal bilayers on a sapphire (Al$_{2}$O$_{3}$) substrate (11.0 orientation) with a Nb buffer \cite{WardPCM2008}. We have then exposed these samples to a hydrogen containing gas whilst probing the structure with synchrotron X-rays. This experimental arrangement has the advantages of a very clean, nominally epitaxial uranium, well-defined model structure, from which certain reflections (both in the metal and oxide) can be followed with precision as the hydrogen interacts with the sample.

The areas of greatest uncertainty in our understanding relate to the earliest part of the reaction, i.e. when hydrogen traverses the oxide, enters the metal, builds to some terminal concentration at a given point and nucleates as UH$_{3}$. We build on previous work with coupon samples and follow intensities and lattice parameters of the metal and oxide of our model system. Our observations from the synchrotron experiments are supported by Scanning Transmission Electron Microscopy (STEM) and Electron Energy-Loss Spectroscopy (EELS) measurements on additional samples exposed to hydrogen.

\section{Experimental Details}
\subsection{Sample preparation}
Samples of three types were prepared using the method of DC magnetron sputtering at the University of Bristol according to a method described elsewhere \cite{WardPCM2008}. The U metal layers of between 60 and 100 nm were deposited on a Nb buffer at 600 C, giving a predominantly U(110) orientation epitaxial film \cite{WardPCM2008}.  Differing oxide overlayers were produced through either depositing an oxide through reactive sputter deposition or leaving the metal to oxidise in air. In the deposited case, by DC magnetron reactive sputtering, an oxide thickness of $\sim$ 20nm was created at 7 x 10$^{-3}$ mbar Ar/2 x 10$^{-5}$ mbar O$_{2}$, 25-30 C on top of the uranium metal. In the air-grown case, the oxide overlayer was allowed to develop by exposure to laboratory air at room temperature over two days. A sample was also produced with no intentional oxide grown where the prepared uranium thin film was transferred to the vacuum system with a few minutes. This ‘uncapped’ sample was exposed to hydrogen and used for STEM analysis. A table detailing the samples depicted in each figure is contained in the appendix. %A sample was also produced with no intentional oxide grown as the sample was transferred in a few minutes to the vacuum system.

The Nb and U are deposited at high temperature to allow epitaxy and this results in a uranium film exhibiting predominantly $\alpha$-U (110) crystallites, with a minor contribution from those oriented with the $\alpha$-U (002) planes \cite{WardPCM2008}. The crystallites of $\alpha$-U (110) orientation are probably as thick as the film (60 nm) but the $\alpha$-U (002) crystallites may extend no more than 20 nm in the growth direction.  Importantly, there is no epitaxy between the uranium metal and the oxide.  %Uranium air-grown oxide on a prepared polycrystalline uranium metal surface is known to be nanocrystalline with an anticipated crystallite size of $\sim$12 nm \cite{Jones2008}. The air-grown oxide grown on bulk polycrystalline uranium is strongly textured in the (110) direction, and give both the (111) and (200) reflections \cite{Chernia2006}. The most intense UO$_{2}$ reflection would be the (220), but at 2$\theta$ = 47.25$^{\circ}$ this was beyond the range of our observations.

\subsection{Synchrotron experimental configuration}
XMaS (Beamline BM28) is a bending magnet beamline at the European Synchrotron Radiation Facility (ESRF) and we used photons of incident energy of 8 keV ($\lambda$ = 0.15498 nm) with a 6-circle diffraction-geometry goniometer \cite{XMaS}.

\begin{Figure}
\centering
\includegraphics[width=0.9\textwidth]{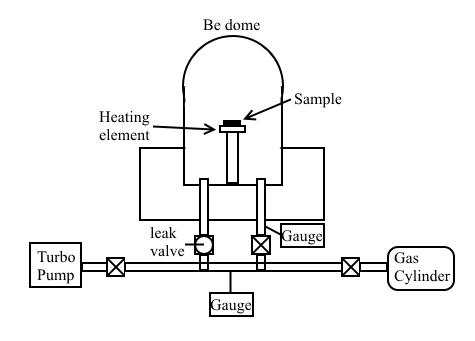}
\captionof{figure}{\label{BeDome} Schematic illustration of the micro-furnace, gas manifold and pumping system used to introduce the H$_{2}$/Ar gas mix to the samples. The micro-furnace was situated on the goniometer allowing an in-situ experiment.}
\end{Figure}

\begin{figure*}
\centering
\includegraphics[width=0.95\textwidth]{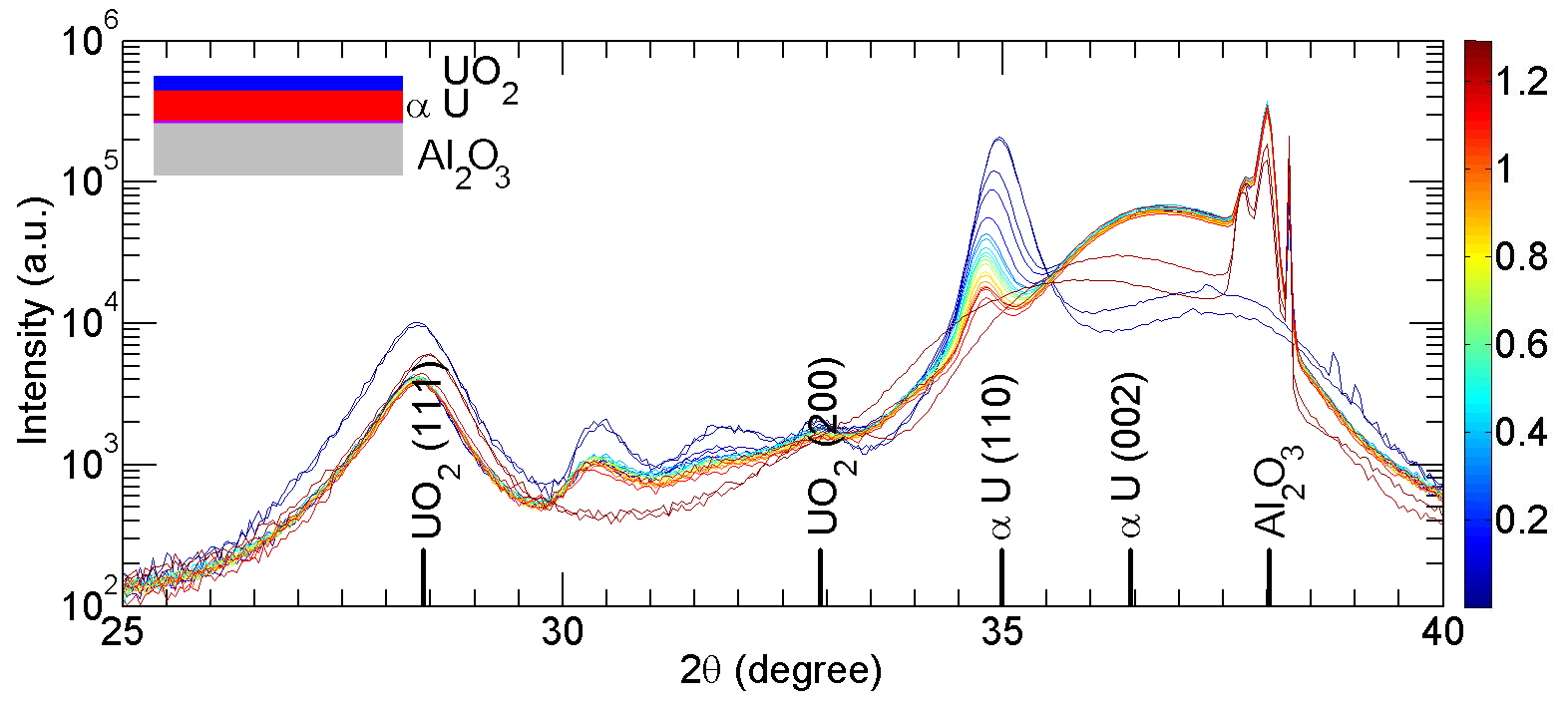}
\caption{\label{SN782_TT} For the as-deposited oxide sample, a plot of intensity vs 2$\theta$ with the line colour representing exposure to hydrogen dose of the high-angle x-ray diffraction taken with increasing cumulative dose. The colour coding indicated on the right-hand side corresponds to the total accumulative dose (min.bar). The nominal bilayer thickness was 60 nm of U covered by $\sim$ 30 nm UO$_{2}$. The inset in the upper left is a schematic of the sample.}
\end{figure*}

Samples for the XRD experiments were loaded into a gas-cell consisting of a resistively heated sample stage (controlled remotely via thermocouples attached to the stage and a Proportional Integral Derivative controller) within an X-ray transparent beryllium dome. The cell could be evacuated to a base pressure of 10$^{-3}$ mbar and refilled with gas to 1000 mbar. Gas was supplied and extracted through a gas manifold attached to the cell. This allowed manual delivery of 4$\%$H$_{2}$ in Ar (N5.5 purity); cell pressure could be monitored by means of pressure gauges. All subsequent reporting of pressure within this paper relates to the partial pressure of H$_{2}$ (P$_{H_{2}}$) delivered. A schematic of the arrangement is shown in Figure \ref{BeDome}. XRD patterns were characterised by very intense reflections from the substrate (Al$_{2}$O$_{3}$) at 2$\theta$ = 38.02$^{\circ}$. Samples were always aligned to maximize the $\alpha$-U(110) peak using a omega rocking curve leading to small changes in alignment during the experiment and a change of intensity from the narrow substrate (Al$_{2}$O$_{3}$) reflection.

Prior to hydrogen exposure samples were subject to a vacuum thermal pre-treatment (12.6 hrs in the case of the deposited oxide and 1.5 hrs for the air grown sample) at 200 C/10$^{-3}$ mbar during which X-ray spectra were largely unchanged. %The deposited oxide sample was subjected to a number of intermittent hydrogen doses (approximately 2.4 hrs at 2.5 mbar PH$_{2}$/ 80 C), followed by a long dose (8.5 hrs at 2 mbar PH$_{2}$/80 C), followed by a long dose (1 hr at 1.1 mbar PH$_{2}$/200 C). The air grown oxide sample was subjected to a single long dose (11 hrs at 400 mbar PH$_{2}$/80$^{\circ}$ then 140$^{\circ}$ then 200 C).

Due to the range of doses and temperatures explored throughout this investigation it is pertinent to compare them via a product of pressure and time. Therefore we have utilised the term dose to refer to the product of the partial pressure of hydrogen and the time of exposure (units min.bar). We are aware that an exact equivalence of the factors of time and pressure may not be wholly justified, especially in the case of saturation effects, but gives a simple definition of dose that may be used to compare experiments. In the two samples examined with synchrotron radiation, there were also intentional changes in temperature during the hydrogen exposure: these are indicated on later graphs with vertical lines.

\subsection{Atomic Resolution Microscope}
The TEM foil (of approximate dimensions 20x10x0.02 $\mu$m) was produced via Focused Ion Beam (FIB) using a Pt protective layer to protect the sample during milling and thinning. The foils were prepared on a FEI Helios Nanolab 600i dual-beam at Bristol before being transported to Oxford for STEM and EELS measurements. STEM and EELS was undertaken at the University of Oxford with a JOEL ARM 3000F. Measurements taken at 200 kV allowed EELS spectra to be obtained simultaneously with high-annular imaging. All EELS data are calibrated in energy and normalised for thickness using the zero-loss peak obtained as part of, or simultaneously with, the spectra of interest. The energy resolution of the EELS system is determined via the full width at half maximum of the zero-loss peak and is of the order of $\sim$0.6 eV.

\section{Results}
The results section is organised in terms of experiments carried out at the ESRF on an as-deposited oxide sample and an air grown oxide sample. STEM and EELS experiments on further samples not exposed in-situ then follow. A full-width at half maximum (FWHM) analysis of the two samples used for diffraction is included just before the Discussion.
\subsection{Deposited Oxide}
The first series of experiments were conducted on a bilayer of $\alpha$-uranium with a reactively sputtered uranium dioxide overlayer deposited at room temperature, thus producing an oxide from pure oxygen and without consumption of the metal. There is no epitaxy at the oxide/metal interface.

%The high-angle x-ray diffraction measurements (Figure \ref{SN782_TT}) of the bilayer contains a strong contribution from the substrate, which in this work is Al$_{2}$O$_{3}$ (sapphire) with the strongest peak (110) at 2$\theta$ = 38.02$^{\circ}$. As this is not the focus of the work, each scan was aligned to maximize the $\alpha$-U (110) peak using a omega rocking curve, and minimise the contribution from the substrate, which has a very narrow rocking curve. (i.e in the direction perpendicular to the specular scan shown in Figure \ref{SN782_TT}). %This leads to small changes in alignment during the experiment and changes of intensity from the sapphire reflection, and the associated reflectivity oscillations that arise from the bilayer and buffer. The observation of reflectivity oscillations throughout this experiment proves that the interfaces are always well defined and smooth, which is different in the second sample discussed below.
The XRD pattern (Figure \ref{SN782_TT}) is characterised by strong reflections of the $\alpha$-U (110) at 2$\theta$ = 35.2$^{\circ}$ and that of UO$_{2}$ (111) at 2$\theta$ = 28.5$^{\circ}$. In addition to these reflections, there is broad intensity around the $\alpha$-U (002), but a quantitative analysis of this is prevented by the interference with intensity oscillations from thickness fringes from the buffer and metal layer. Finally, there are two minor reflections at $\sim$30.5$^{\circ}$, and 32$^{\circ}$ (both of which change with exposure to hydrogen), and a weak contribution from the UO$_{2}$ (200).

Upon exposure to hydrogen the intensity of the major $\alpha$-U (110) reflection is steadily decreased over the 80 C hydrogen exposures. For the final two scans the temperature was raised to 200 C, but by this time the reflections from the metal were very weak.
In as-deposited UO$_{2}$ films, we find a strong $\left\langle111\right\rangle$ texturing, which is different from that found in air-grown films (see following section), so the dominance in this case of the UO$_{2}$ (111) over the UO$_{2}$ (200) is not surprising. The weak reflection at $\sim$30.5$^{\circ}$ is the $\alpha$-U (020), which is calculated at 30.7$^{\circ}$, and this is consumed by hydrogen dosing, as expected. The weak intensity and the breadth of the peak suggest only very small amounts are present. 

\begin{Figure}
\centering
\includegraphics[width=0.9\textwidth]{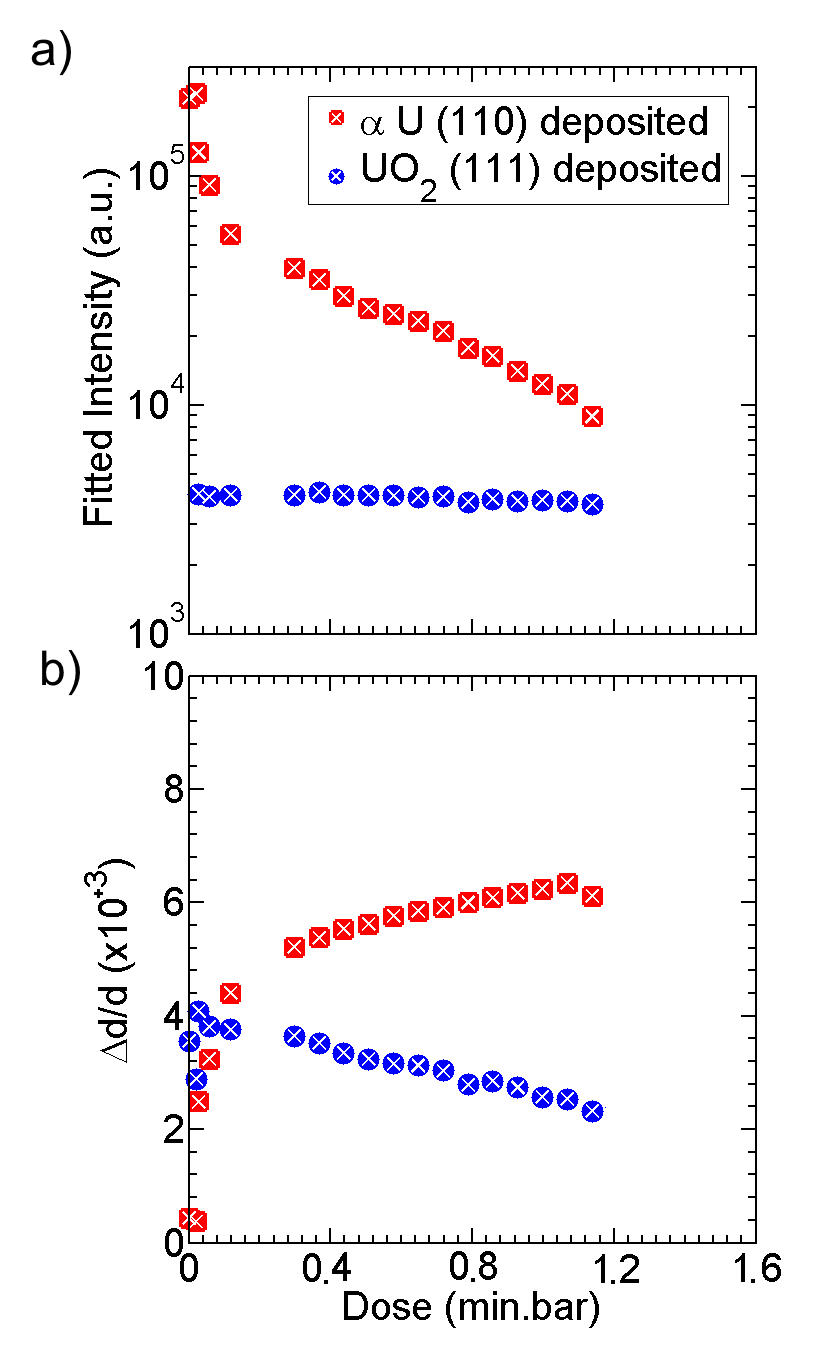}
\captionof{figure}{\label{SN782IntDd}a) Intensities and b) peak positions (converted into changes in d-spacing, where $\Delta$d = 0 indicates the bulk value) as a function of hydrogen dose.  All doses were given with the sample temperature 80 C.  At 25 C the tabulated bulk d-spacings of $\alpha$-U (110) = 0.2566 nm and of UO$_{2}$ (111) = 0.3158 nm.}
\end{Figure}

The peak at 32$^{\circ}$ is related to the formation of UO$_{2+x}$ \cite{Rousseau2006} and implies that cubic $\left\langle200\right\rangle$ type reflections in UO$_{2}$ start to split as the tetragonal nature of U$_{3}$O$_{7}$ (x = 0.33) makes reflections separated by $\sim$1$^{\circ}$ appear (at this wavelength). As expected, these reflections disappear with hydrogen dosing, as the UO$_{2}$ is reduced to stoichiometry, and thus cubic. There are no indications of in-growths of reflections attributable to either of the polymorphs of UH$_{3}$; the strongest reflection of the expected $\beta$-UH$_{3}$ (210) would be observable at 2$\theta$ = 30.2$^{\circ}$.

Our main aim is to consider the $\alpha$-U (110) and UO$_{2}$ (111) reflections as a function of dosing. We do this by analysing the intensities and positions of these reflections and presenting the results in Figure \ref{SN782IntDd}. The behaviour of the full-widths at half maximum (FWHM) with dose of all the reflections will be discussed later due to the association with crystallite size, section \ref{FWHMsec}.

The drop in intensity of the $\alpha$-U (110) after the first point almost certainly arises from a change in alignment. However, the initial increase of $\Delta$d/d in both materials is close to that expected from thermal expansion when heating from 25 to 80 C \cite{Martin1988, Lloyd1966}. %The U metal also expands as the H$_{2}$ gas enters the metal.
In contrast, the UO$_{2}$ (111) does not change in intensity, but the lattice contracts towards the bulk value. The opposite behaviour of the U and UO$_{2}$ lattices strongly increases the strain at the smooth interface between these two layers.% and it is at this interface where we assume that the UH$_{3}$ is formed. %We know the interface is smooth as reflectivity curves were obtained for all doses in this sample.

\begin{figure*}
\centering
\includegraphics[width=0.95\textwidth]{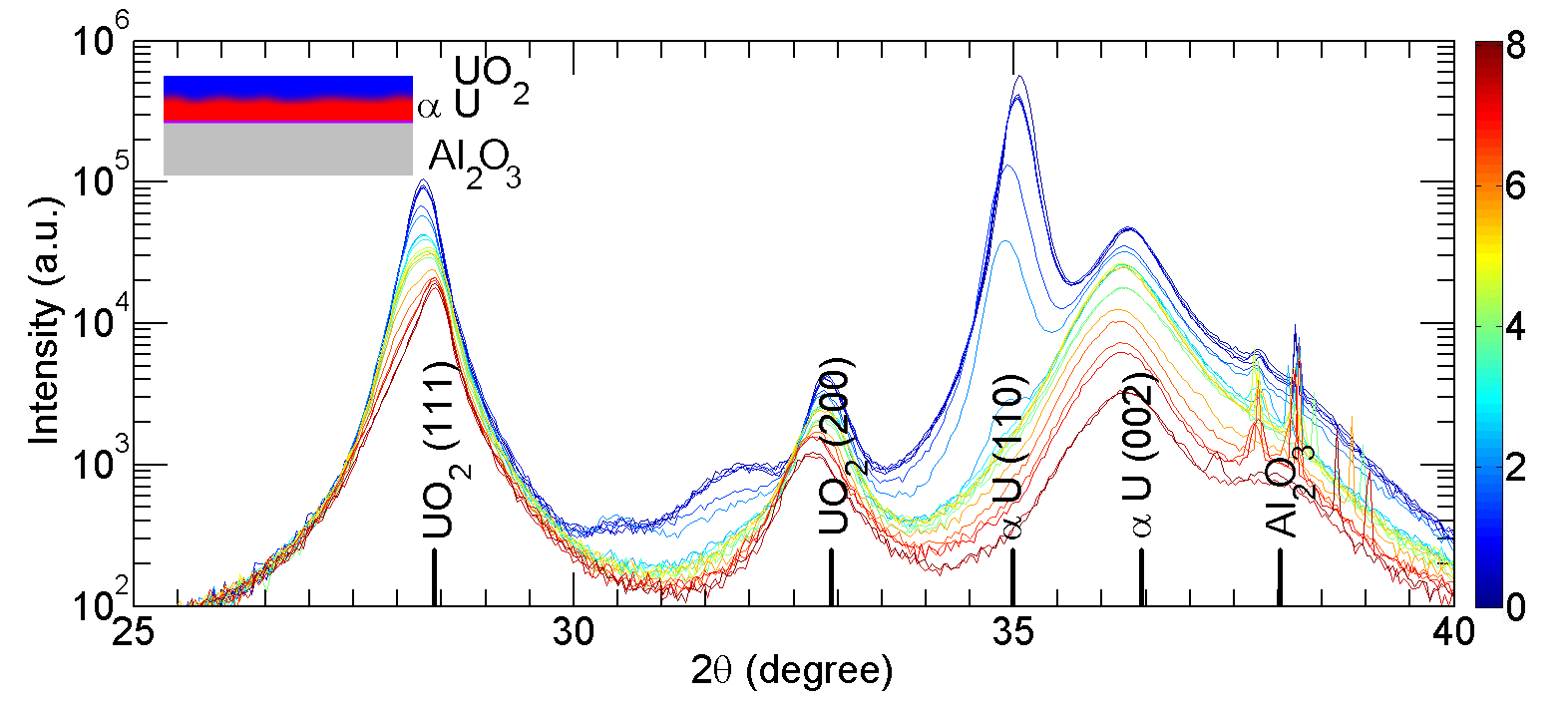}
\caption{\label{SN790_TT} For the air-grown oxide sample. a plot of intensity vs 2$\theta$ with the line colour representing exposure to hydrogen dose of the high-angle x-ray diffraction taken with increasing cumulative dose. The colour coding indicated on the right-hand side corresponds to the total accumulative dose (min.bar). Before exposure to air the nominal thickness of U metal was $\sim$100 nm. After reaction with air, we estimate that the U metal reduced to 80 nm, with some $\sim$60 nm of oxide. The oxide/metal interface is not smooth. The inset in the upper left is a schematic of the sample.
}
\end{figure*}

\subsection{Air grown oxide}
Oxide overlayers were allowed to grow naturally, by removing the $\alpha$-uranium sample from the sputtering machine and leaving it in air for 2 days at room temperature, in this case oxide growth is the result of consumption of uranium metal.

The air-grown oxide sample is fundamentally different to that discussed above for the as-deposited sample. The oxide-metal interface is now not smooth, as it is the result of a conversion of the metal, rather than a layer added on top. %This means that the reflectivity signal from the sample at angles of more than $\sim$ 5o drops to almost zero, and the oscillatory behaviour of the fringes observed in Figure 2 is no longer present. Likewise the fringes associated with interference with the strong sapphire reflection are also much reduced. 

The rough oxide/metal interface reduces dramatically the reflectivity, and the intensity oscillations associated with the substrate and buffer. This allows a much better observation of the $\alpha$-U (002) reflection in the air-grown oxide sample (Figure 4), which was not possible with the as-deposited oxide sample (Figure 2). %The oxide exhibits two reflections possibly because the texturing of the air-grown oxide layer is (110) \cite{Chernia2006}, so both (111) and (200) oxide reflections can be observed. 
The reflection near 32$^{\circ}$ is associated with hyperstoichiometric UO$_{2+x}$ and is removed as the oxide is reduced in stoichiometry by hydrogen exposure (as was seen for the as-deposited oxide sample, see Figure 2).

\begin{figure*}
\centering
\includegraphics[width=0.7\textwidth]{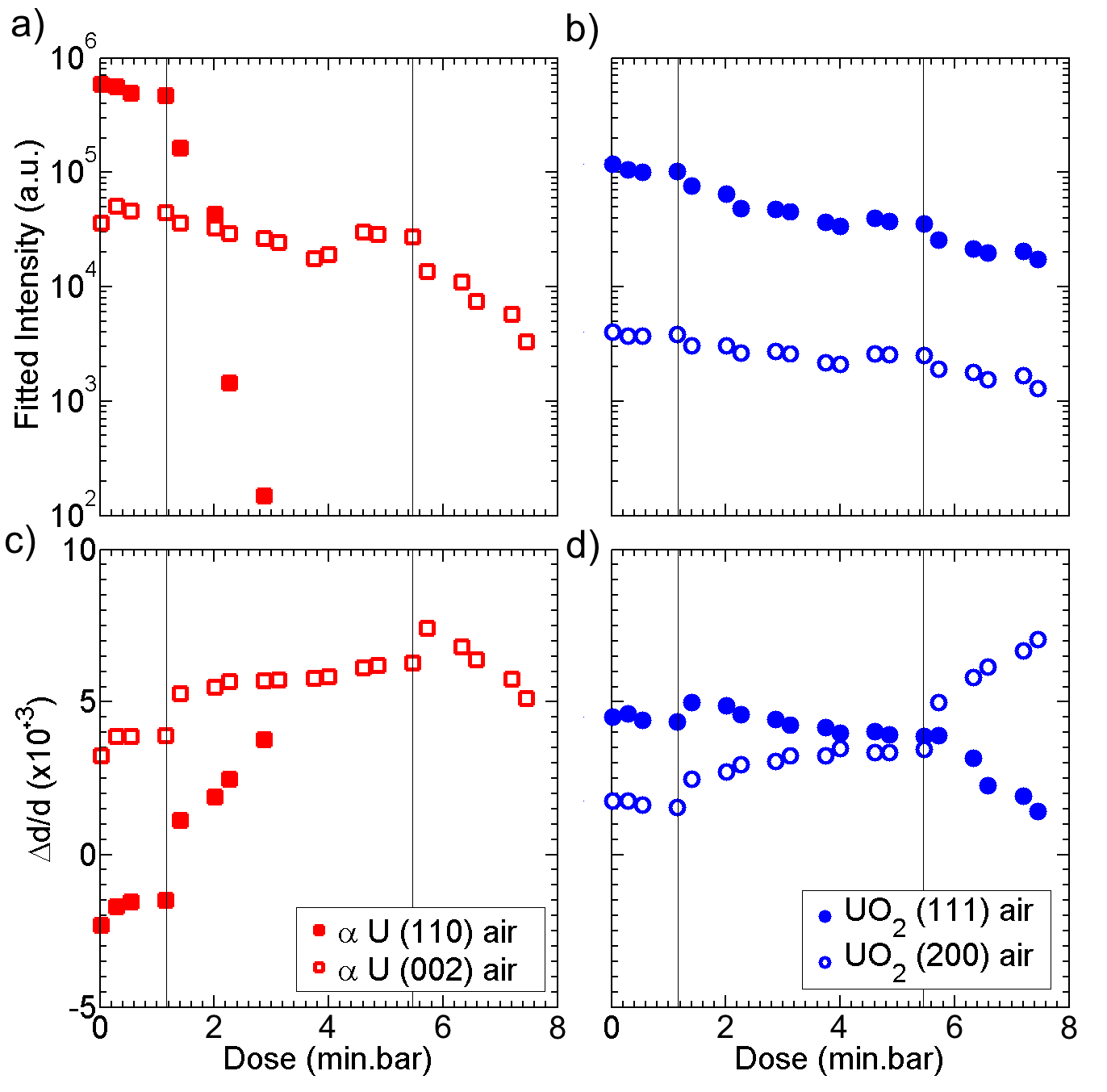}
\caption{\label{SN790IntDd} a) Intensities and b) $\Delta$d/d as a function of hydrogen dose.  Initially, the doses were given with the sample at 80 C, but the temperature was increased to 140 C at the first vertical line, and then to 200 C at the 2nd vertical line. At 25 C the tabulated bulk d-spacings of $\alpha$-U (002) = 0.2478 nm and of UO$_{2}$ (200) = 0.2735 nm. }
\end{figure*}

Plotting the intensity and $\Delta$d/d in Figure \ref{SN790IntDd} for all major reflections illustrates differences in the effect of H$_{2}$ dosing on different crystallographic orientations.

With exposure to H$_{2}$ gas there is clearly a difference in the consumption of the metal depending on the orientation. The required dose to consume the uranium metal is also greater in the air-grown oxide case than the deposited oxide case. To increase the reactivity and thus kinetics of the dose the temperature was raised, first to 140, and then to 200 C. The vertical lines in Figure \ref{SN790IntDd} indicate where the temperature was raised. The dose that resulted in complete consumption of uranium for the as-deposited oxide sample (1.2 min.bar) had little effect on the intensities or positions of metal or oxide Bragg peaks for the air-grown oxide sample. This suggests the air-grown oxide layer has formed a protective layer over the metal resulting in a lower rate of consumption of crystalline uranium via reaction with H$_{2}$. The increase of $\Delta$d/d for both U-metal and oxide peaks after the 1st point is due to thermal expansion \cite{Martin1988, Lloyd1966}, but after that both d-spacings show no change; no H$_{2}$ is entering the metal lattice. However, when the sample temperature is raised from 80 to 140 C (at first vertical line in Figure \ref{SN790IntDd}), a dramatic change takes place, with rapid expansion of the U lattice, and consumption of metal. A strong orientational dependence is observed; with the $\alpha$-U (110) Bragg peaks decreasing much faster in intensity than the $\alpha$-U (002) as is clear in Figure \ref{SN790_TT}.

As discussed above, for the as-deposited oxide sample there was no change in UO$_{2}$(111) intensity over the experiment. However, as shown in Figure 5, for the air-grown oxide sample a decrease in intensity was observed over the entire experiment (also observable in Figure 4). The intensities of both UO$_{2}$(111) and UO$_{2}$(200) start to decrease at around 1.2 min.bar when the temperature is increased to 140 C. Also at this increased temperature, the respective d-spacings show different behaviour: the UO$_{2}$(111) decreasing (after an initial increase due to the change in temperature) and the UO$_{2}$(200) increasing. Later in the experiment at around 5.6 min.bar, when the temperature is further increased to 200 C, the changes in these two d-spacings are further accentuated with a more rapid decrease and increase in $\Delta$d/d for UO$_{2}$(111) and UO$_{2}$(200) respectively.

It should be pointed out that such behaviour in one single crystallite cannot occur, but the diffraction experiments along the specular direction (as reported here) probe only crystallites with their respective axes parallel to the overall film growth direction, so these results come from different crystallites, and are subject to different strain.

\subsection{Scanning Transmission Electron Microscopy}

A third type of sample, without any intentionally deposited or grown oxide (i.e. uncapped), was exposed to hydrogen and examined with STEM. This uncapped sample of $\sim$45 nm was subjected to 0.5 bar of pure hydrogen at 140 C for 20 minutes before the hydrogen was removed and the sample was allowed to cool. With this dose level (10 min.bar) blisters were formed and the blister example shown in Figure 6a) was observed to be coincident with a crack through the uranium film. It is not clear whether the presence of a pre-existing crack caused the precipitation here or whether the precipitation of low density corrosion product caused the crack in the uranium metal. At the edges, where the blister is expanding laterally, the hydride is not crystalline suggesting that under some circumstances its formation will not give rise to sharp diffraction peaks. It is also noticeable from this figure that the blister appears to have formed at the interface between the uranium and the buffer layer, rather than at the free surface. This suggests that to form hydride requires a concentration of hydrogen in uranium and a nucleation point such as a defect. However this sample has seen a high hydrogen dose and recall that this sample has no (or very little) oxide, so that it is not representative of the earlier samples discussed above. %e can anticipate that the oxide/metal interface contains many defects and points for potential hydride formation. 

\begin{figure*}
\centering
\includegraphics[width=0.95\textwidth]{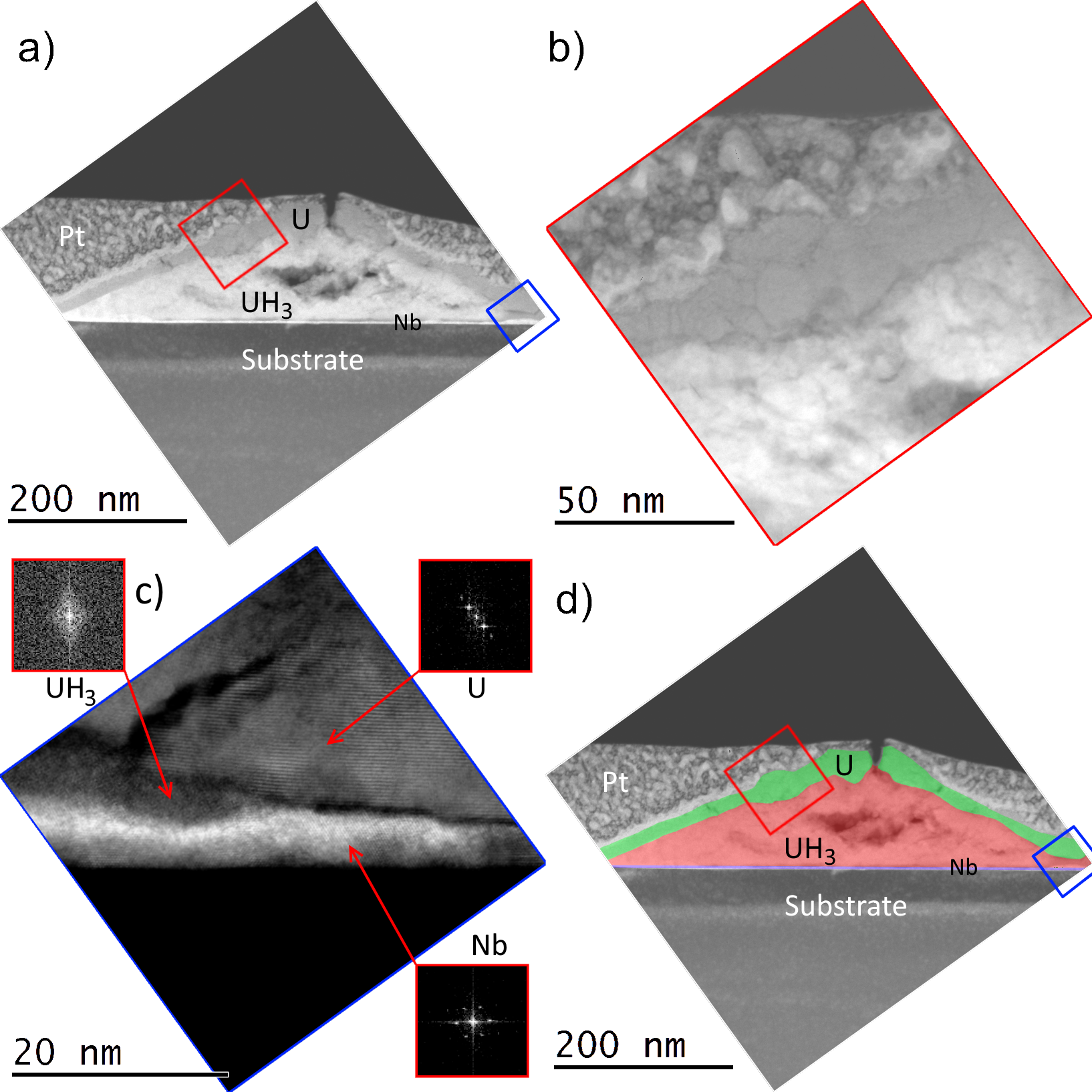}
\caption{\label{TEM} High Resolution STEM Annular Dark Field Images showing a blister of lower density material (UH$_{3}$ is inferred) at the interface between the uranium-metal thin film and the buffer layer. a) Overall micrograph showing this precipitation location is coincident with a crack in the metal thin film. b) Region selected from centre left of a) showing remaining uranium thin film ($\sim$35 nm at thickest point). c) Region selected from extreme right of a) demonstrating that hydride formation close to the metal-hydride interface is essentially amorphous. Included in c) are Fast Fourier Transforms (FFT) to illustrate the crystallinity of the diffraction patterns of hydride (amorphous), U metal and Nb interlayer (both crystalline). d) Overall micrograph with false colour to highlight U (green), UH3 (red) and Nb-buffer layer (blue). The TEM foil was produced via Focused Ion Beam (FIB) using a Pt protective layer to protect the sample during milling and thinning.}
\end{figure*}

\subsection{Electron Energy Loss Spectroscopy}

\begin{figure*}
\centering
\includegraphics[width=0.95\textwidth]{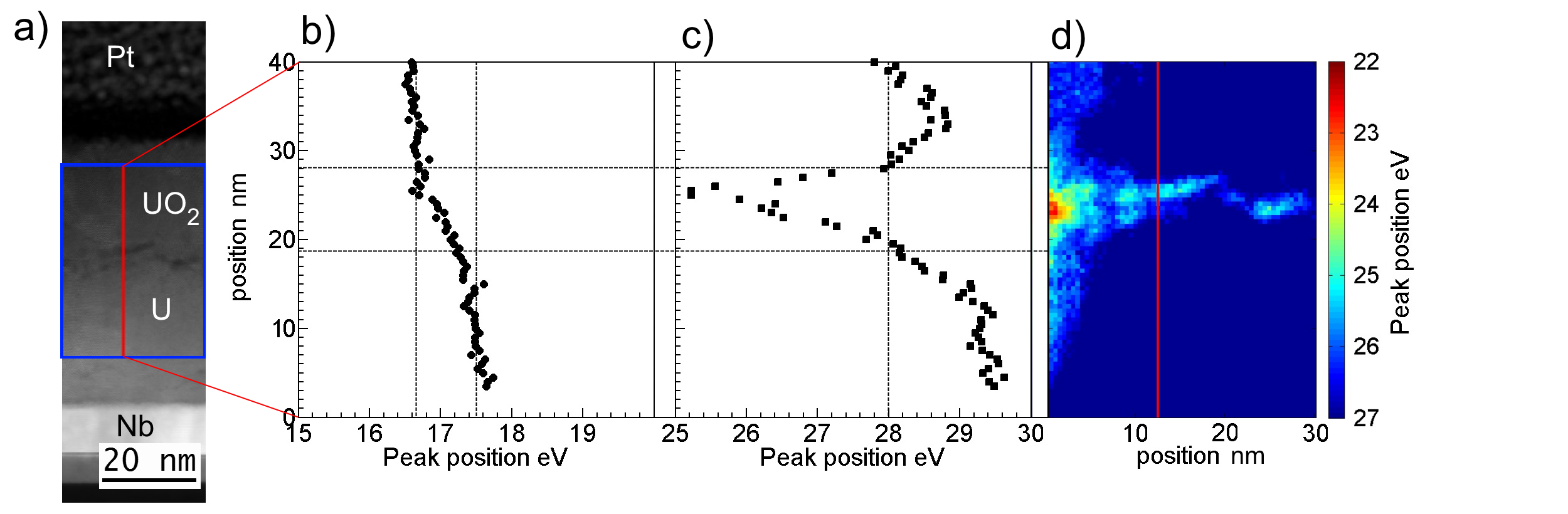}
\caption{\label{EELS} a) STEM Annular Dark Field image taken simultaneously with EELS over a 40 nm location range spanning the U-UO$_{2}$ interface. Analysis of the uranium b) P3 and c) P2 edge energy positions highlights three distinct regions: U (between locations 0 and 19 nm), UO$_{2}$ (between locations 28 and 40 nm) and UH3 (between locations 19 and 28 nm). d) A colour map of the blue region from a) showing areas where the P2 edge has been shifted to lower energy suggesting uranium bonding to hydrogen.}
\end{figure*}

Hydriding an as-deposited oxide/uranium bilayer at 80 C in 0.5 bar of 4$\%$ H$_{2}$ mix for 30 mins (0.6 min.bar equivalent dose) does not show complete consumption of the uranium metal and gives an opportunity to observe where early hydride precipitation occurs, Figure \ref{EELS}. Conducting EELS collection over an energy region of 10-60 eV interrogates the P2 and P3 absortion edges which equate to the 6\emph{p} to 6\emph{d} transitions in uranium. 

Fitting the edge positions for every pixel collected can help to build up a picture of the uranium bonding in different regions of the sample. For clarity the fitted peak positions are shown Figure \ref{EELS}b-c) for a line profile, shown in Figure \ref{EELS}a), across the interface between the metal and the oxide. The peak position of the P3 edge, Figure \ref{EELS}b), shifting from 16.6eV (oxide) to 17.5eV (metal) shows a distinct difference in the environment of the uranium metallic and ceramic and is in line with the observation of others \cite{DEGUELDRE2013, DEGUELDRE2015}. This is mimicked in the uranium O4,5 edges measured at 111.4eV (oxide) and 112.2eV (metal) \cite{Moore2007}. There is also a difference in the P2 edge peak position, Figure \ref{EELS}c), between oxide (28-28.3eV) and metal (29.2eV) with a large shift to a distinct lower energy at the interface between the metal and the oxide indicating a different uranium environment and possible valence state. Mapping the regions where the fitted peak position of the P2 edge is less than 27eV, and so neither oxide nor metal results in highlighting the interface region shown in Figure \ref{EELS}d). In summary, after only a low dose of H$_{2}$ at 80 C (0.6 min.bar) we have good evidence for a distinctly different uranium environment (perhaps a different compound) at the interface region. This would suggest that our first UH$_{3}$ precipitation (initiation) is in this interface region between the U and UO$_{2}$, where there are many defects.

\section{Analysis of full-width half maximum of diffraction peaks}\label{FWHMsec}

\begin{figure*}
\centering
\includegraphics[width=0.95\textwidth]{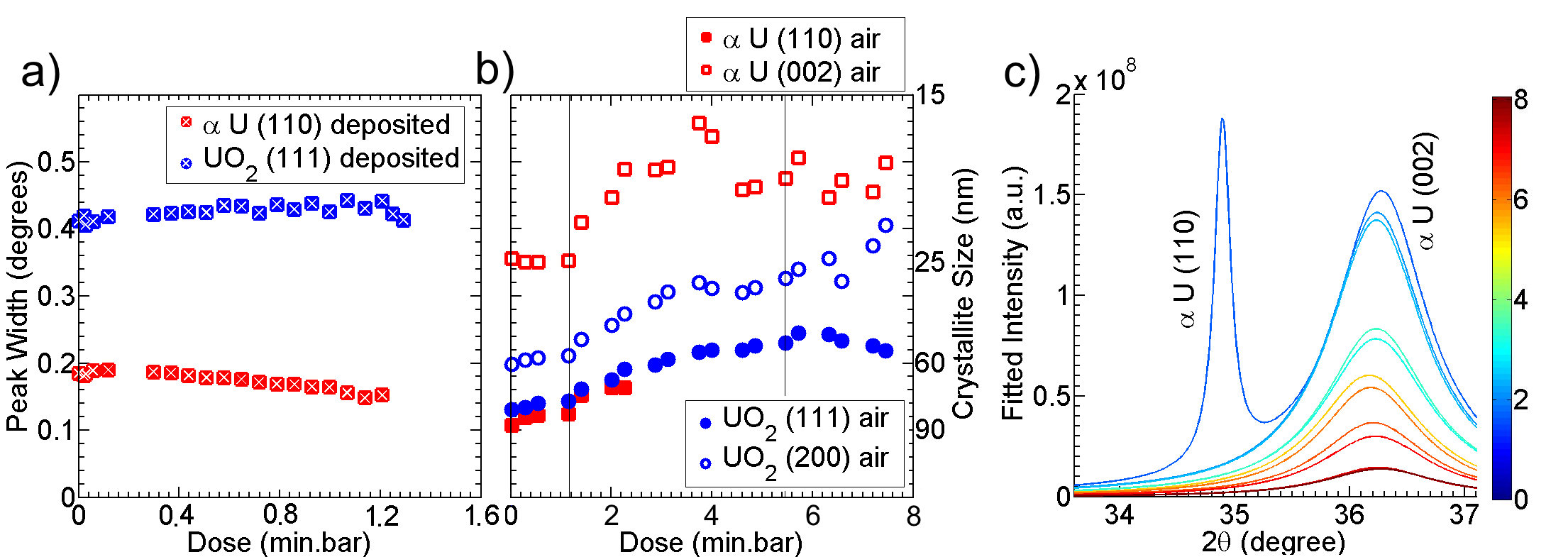}
\caption{\label{Scherrer} FWHM of the diffraction peaks of the (a) as-deposited oxide and (b) air-grown oxide samples. Error bars are small on frame (a) but increase to ~0.1 degree for the wider reflections on frame (b). Calculated crystallite size using the Scherrer equation for both samples is included on the right-hand-side of b). c) shows an example of the quality of the data from the air-grown sample (on a linear scale) used to determine the FWHM, and demonstrates clearly the much greater width of the $\alpha$-U (002) as compared with the $\alpha$-U (110).}
\end{figure*}

The full-width at half maximum (FWHM) is directly related to the size of the crystallites giving rise to the diffraction peak. The specular geometry of the experiment dictates the probing direction is perpendicular to the bilayer surface, i.e. in the growth direction. This crystallite size is limited by the bilayer dimension of $\sim$100 nm, leading to the FWHM being much broader than the instrumental resolution. In this case we have an experimental determination of the resolution width from the substrate sapphire peak, which has a FWHM =0.02$^{\circ}$. Normally one should perform a convolution of the measured FWHM with the instrumental resolution to obtain the observed FWHM, but in this case the peaks from the bilayer are so much wider than the resolution, that they may be used directly as the observed values. The FWHM of the major U and UO$_{2}$ reflections for the as-deposited and air grown oxide samples observed by the X-ray experiments, as a function of H$_{2}$ dose, are shown in Figure 8.

Using the Scherrer formula:
\begin{equation}
\tau = \frac{\kappa  \lambda}{ (FWHM)  cos(\theta)}
\label{eq:Scherrer}
\end{equation}

where $\tau$ is the coherence length, $\kappa$ is a shape factor, which we take as unity, and $\theta$ is the Bragg angle, we determine the relationship between FWHM and crystallite size. This relationship is shown on the right-hand side of the center panel of Figure \ref{Scherrer}. These are dimensions in the growth direction. We are less concerned here with the absolute values, but more with changes as a function of H$_{2}$ dosing. Since a number of different mechanisms can contribute to the width of a reflection, for example inhomogeneous strain across the crystallite, the Scherrer formula gives a lower bound to the crystallite size. The values of crystallite size derived from the Scherrer equation are broadly in line or smaller than the film thicknesses themselves justifying our use of $\kappa$ = 1.

For the as-deposited oxide, there is little change in the FWHM as a function of H$_{2}$ dose, and the widths correspond approximately to the nominal ones of 60 nm U metal, and 30 nm of dioxide. This absence of any broadening as the uranium is consumed (in both samples) puts important constraints on the mechanism of uranium conversion to hydride.
However, for the air-grown oxide there are profound effects when the second stage (at T = 140 C) dosing occurs. For both the $\alpha$-U(110) and $\alpha$-U(002) grains the crystallite size is relatively unchanged during the 80 C part of the experiment. For the subsequent part of the experiment at 140 C the uranium is consumed steadily and the crystallite size then decreases by some 20-30$\%$ up to a dose of $\sim$ 2.5 min.bar.  %and we can associate this to their break up, which causes a vast increase in the rate of hydrogen entering the U metal lattice (see Figure \ref{SN790IntDd}), and no doubt their physical breakup is associated with a large increase in the interfacial UO$_{2}$/U metal strain as well as possible sublimation of hydroxyl radicals.

\section{Discussion}
\subsection{Changes in the uranium metal}

Intensity from uranium metal reflections decreases substantially in the diffraction patterns in both samples (Figures \ref{SN782_TT} and \ref{SN790_TT}) during hydrogen exposure. Since no change occurs during the annealing treatment before hydriding, this change can be attributed to hydrogen exposure. The absence of diffraction lines from the UH$_{3}$ will be discussed later, but is related to the amorphous and/or nanocrystalline form found by the TEM investigation shown in Figure 6. 

The reduction in metal intensity of $\alpha$-U(110) is different for the as-deposited and air-grown samples, as shown in Figure 9a), where we have plotted the normalised intensities of the two samples as a function of dose. The rate of removal of $\alpha$-U(110) in the as-deposited sample (80 C) would appear to follow a parabolic relationship with time where the rate of removal is initially fast, slowing as the experiment progresses (Figure 9a). If the intensity is plotted against log of the dose, Figure 9b), then we should expect a straight line with the slope related to the rate, as is observed. This form of kinetics is consistent with the corrosion product forming an adherent product layer which itself throttles hydrogen transport to the metal; in such a scenario the growing UH$_{3}$ layer itself acts as a barrier to hydrogen flux and slows the hydriding reaction. This explanation firstly requires product formation between the source of hydrogen and the uranium metal (it is generally accepted in the literature that this occurs at the UO$_{2}$-U interface) and, secondly, a non-spalling UH$_{3}$ product layer. Alire et al \cite{RH1} have reported the effect of adherent hydride layers on initial take-up of hydrogen with an initial rapid rate followed by a reduction in rate as an adherent layer grows in thickness. Based on the quantity of uranium consumed in their experiments in order to reach a steady, initial minimum reaction rate, we could estimate that Alire had an adherent layer of UD$_{3}$ of $\sim$10 $\mu$m at 175 C, i.e. more than two orders of magnitude larger than our films ($\sim$30 nm). It is thus likely that the hydride product in both samples discussed here will be adherent and therefore will provide a barrier to slow the hydriding rate.

\begin{Figure}
\centering
\includegraphics[width=0.9\textwidth]{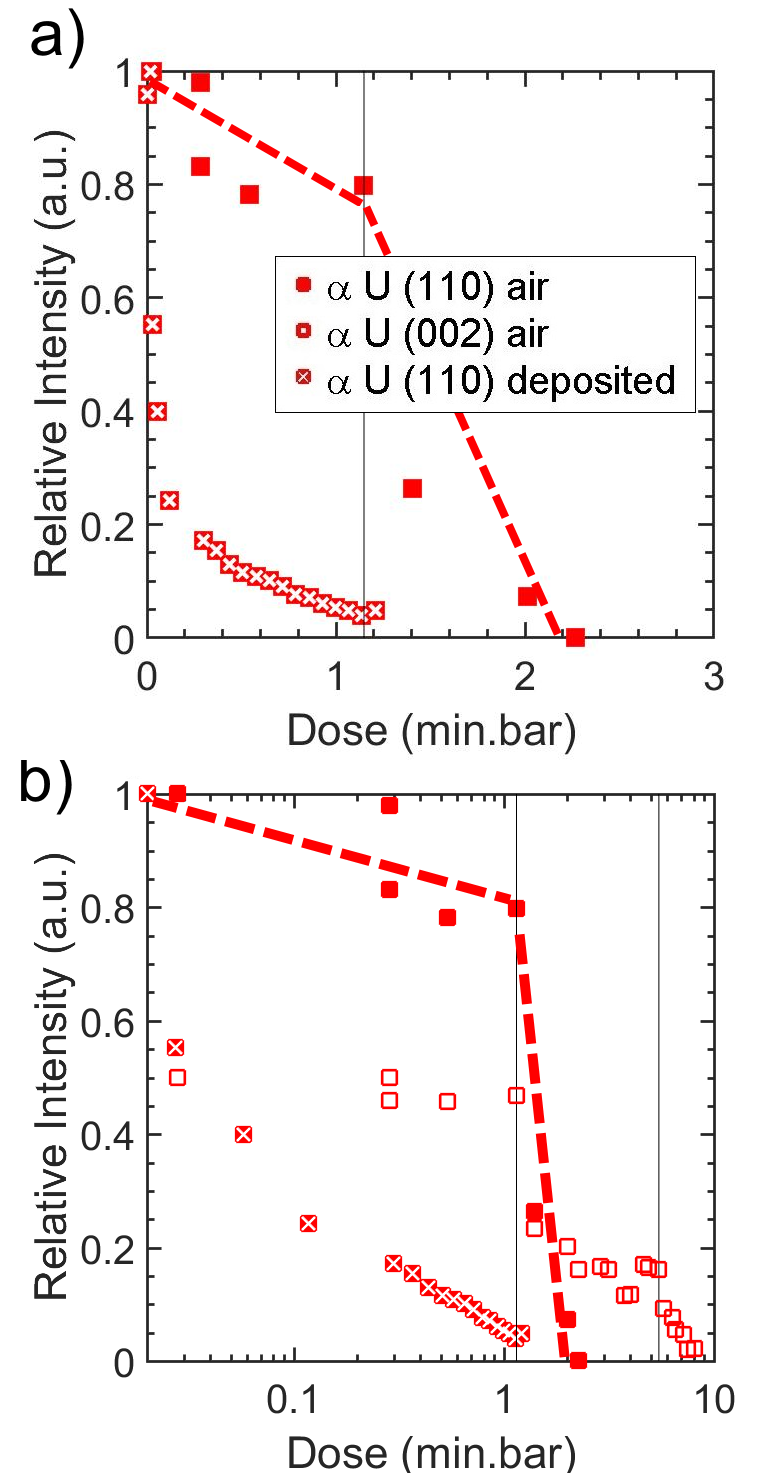}
\captionof{figure}{\label{Combined} Relative intensity of $\alpha$-U(110) and $\alpha$-U(002) plotted as a function of a) linear dose, and b) log dose. $\alpha$-U(110) have been normalised to intensity = 1 at the start of the experiment whereas, for clarity, $\alpha$-U(002) have been normalised to intensity = 0.5 at zero dose. Lines have been added to the $\alpha$-U(110) data set for the air-grown oxide as a guide to the eye.}
\end{Figure}

For the air-grown sample (Figure 4) during the 80 C portion of the experiment the rate of loss of $\alpha$-U(110) intensity is both less and approximately linear over dose, as shown in Figure 9a). If we remember that the uranium films in the two samples are nominally identical, it becomes apparent that there should be no reason why the air-grown sample would not ultimately form such an adherent barrier: but this type of kinetics is not observed. It appears that, in the case of the air-grown sample, the flux of hydrogen through the air-grown oxide is much lower and this then is the rate-determining step for hydrogen delivery to the UH$_{3}$-U interface; thus the air-grown oxide forms a partial barrier. Presumably at some point in this scenario the UH$_{3}$ layer will thicken to an extent where it will become the rate-limiting barrier and consequently after this point parabolic-like kinetics should take over, and this is indeed observed when the temperature is raised to 140 C.  

We will discuss further the differences between these two oxide layers below; it should be remembered that the oxide overlayer in the air-grown sample is some 3-4 times thicker than that in the as-deposited sample.

Note that if we plot the intensities on a log plot, Figure 9b), the change in intensity of the $\alpha$-U(110) with the air-grown oxide layer at low doses cannot be observed easily, but this figure does emphasise the considerable change that occurs in the hydriding process when the air-grown sample is heated to 140 C, as well as showing the large differences between the hydriding of the $\alpha$-U(110) and $\alpha$-U(002) crystallites. The differences in consumption rates of $\alpha$-U(110) and $\alpha$-U(002) we also observed in a further sample although these data are not shown in the present paper.

For the air-grown oxide sample (Figure 4) we observe that the intensity of the $\alpha$-U(002) is some 10\% of the $\alpha$-U(110), but that the decrease in this reflection with H$_{2}$ dose is far slower than the $\alpha$-U(110), see Figure 9b). 

In considering the differences between these two reflections it is important to consider their respective FWHM values, which are shown expressly in Figure 8. As shown in Figure 8c) on a linear scale the reflections are quite different. The $\alpha$-U(002) intensity is diminished steadily over 80 C, then 140 C, and then 200 C such that it becomes the majority of observed uranium intensity after only a short period at 140 C.

Figure 8 gives information on the vertical (i.e. along the growth direction) size of the crystallites as a function of H$_{2}$ dosing. In all cases the $\alpha$-U(110) FWHM stays essentially constant as a function of dosing, which is quite different to the $\alpha$-U(002) FWHM, and the latter is obviously much smaller than the former, and does not extend over the whole film. We do not have information on the exact spatial position of the $\alpha$-U(002) crystallites, and it could be that they are near the bottom of the film, which might suggest it takes longer for them to be consumed. However, the constancy of the $\alpha$-U(110) FWHM puts constraints on the type of uranium consumption mechanism that can be considered valid.  

Whereas it may be simple to think that the hydriding of uranium occurs here through a lateral hydriding front corroding into the metal grains (i.e. top-down), this does not agree with the evidence, since the crystallite size (probing vertically) does not change markedly during the corrosion. Instead, we must imagine a mechanism where the grains are being corroded without thinning them in a vertical dimension. This evidence then suggests a hypothesis where the uranium is largely consumed from the side, i.e. for instance from a grain boundary. Hydriding of metal at grain boundaries is not a novel idea and examples have been identified in terms of initiation at grain boundaries \cite{Scott2007, Harker2013, Jones2013} and also propagation of reaction fronts in the related plutonium-hydrogen reaction \cite{RH2,RH3}.  

The $\alpha$-U(002) crystallites in the air-grown sample are initially unchanged until a large proportion of the $\alpha$-U(110) is consumed at which point their crystallite size decreases markedly. Given that the oxide is polycrystalline we may assume that the hydrogen may be delivered effectively to many points at the U-UO$_{2}$ interface, as also demonstrated in Figure 7 where hydrogen is inferred along the U-UO$_{2}$ interface, and yet corrosion only takes place at a face perpendicular to the U(110) plane. 

We can also conclude that the $\alpha$-U(002) faces are also apparently less reactive than some of the planes perpendicular to U(110) (Figure 9), but we should remember that their delay in consumption may also be a function of their location in the film. Presumably the factors that govern the propensity of a given face to react to form UH$_{3}$ may be related to how easily that uranium face can accommodate significant amounts of hydrogen and the ease with which those local uranium atoms can rearrange to take up a UH$_{3}$ structure. In addition the reactivity may be inversely related to the hydrogen permeability through that face, because if hydrogen delivered to the U-UH$_{3}$ interface can be readily dispersed into the metal, then it is less likely to build to a high concentration required to precipitate UH$_{3}$.These aspects have been considered by Taylor et al. \cite{RH4}, who assumed that $\alpha$-UH$_{3}$ was the initial product. %Another, rather simpler, possibility is that since the crystallites with the (110) faces are larger than those with (002) faces, the packing is much better for the smaller crystallites, thus hindering the transport of H$_{2}$ to the smaller crystallites. This explanation can be tested in future experiments in which a highly epitaxial sample with the $\alpha$-U(002) direction predominant can be produced \cite{Springell2014}, and similar experiments should then show a rapid reduction of the $\alpha$-U(002) reflection. 
Further information on this anisotropy can be obtained by future experiments in which a highly epitaxial $\alpha$-U (002) sample can be examined \cite{Springell2014}. We have also considered whether there could be a channeling effect. The orthorhombic structure is sufficiently distorted from either bcc or hcp \cite{Axe1994} that such channelling seems very unlikely.

A major question in these studies is related to the location of the initial hydride precipitation. The STEM work (Figure 6) shows that the hydride formation is non-crystalline suggesting why we would not observe diffraction peaks at the initial stages of hydriding. The EELS spectroscopy efforts (Figure 7) have indeed shown that there is a layer different from oxide or U metal at the interface, and it is tempting to assign this to a thin layer of UH$_{3}$, although we have no direct proof. 

We had hoped that some evidence for the spatial formation of the hydride might be obtained from X-ray reflectivity studies. Scattering cross sections in X-ray reflectivity are related to the electronic density (electrons per unit volume), and whereas this is 4.44 e/\AA$^{3}$ for U metal, and 2.64 e/\AA$^{3}$ for UO$_{2}$, the value for the hydride is 2.59 e/\AA$^{3}$, so with X-ray reflectivity the hydride is indistinguishable from the oxide. Neutron reflectivity would be a possibility, provided a large enough sample area can be made, as H and O have quite different scattering amplitudes with neutrons.

The uranium metal reflections are also seen to shift in the diffraction patterns to larger d-spacings (Figures 3b and 5c). For the $\alpha$-U(110) reflection this increase in the d-spacing is around 0.6\% (as-deposited oxide) and 0.65\% (air-grown oxide); for $\alpha$-U(002) the change is slightly less but over a longer dosing period. Whereas it is tempting to explain this increase in d-spacing in terms of increasing concentration of hydrogen in the uranium lattice such an explanation has to be taken with caution. There is a strong correlation between the d-spacing of the uranium metal with the amount of uranium consumed. This could be attributed to a number of effects, the correct answer of which is not clear from these data. If this effect is caused by hydrogen dissolved in the uranium lattice it is in a dilute regime, full saturation is apparently not achieved. %Consideration of Figure 3 in particular shows that there is a direct link between the d-spacing and the fitted intensity of the uranium reflection, i.e. that the distance between the particular atoms being reported by the reflection is related to how much uranium remains to be consumed. Indeed a plot of change in d-spacing against fitted uranium intensity (not shown) results in a convincing relationship. Such a relationship suggests that the change in d-spacing may be mechanically related and that as the film is consumed, the d-spacing of the U(110) planes are increased. Such an observation would suggest that the substrate-Nb-U(110) interfaces remain intact during uranium metal consumption and that increasing pressure from the adherent UH$_{3}$ at the sides of the grains acts to elongate the U(110) grains in the vertical direction. An alternative explanation would be that the uptake of H$_{2}$ by the lattice is small, and we are still within the linear region with no sign of saturation. The majority of H$_{2}$ diffusion would then take place through the grain boundaries.

\subsection{Changes in the uranium oxide overlayers}

There are differences between the as-deposited and air-grown oxide overlayers and this has a profound influence on how the hydriding reaction unfolds. Compared to the strong UO$_{2}$(111) reflection the proportion of UO$_{2}$(200) is around 10\% in the as-deposited oxide (Figure 2), and in air-grown oxide (Figure 4), although it initially appears more, it is in fact much lower at around 4\%. According to Waber et al. \cite{WaberJES1959} and Chernia et al. \cite{Chernia2006} a larger fraction would be expected as the texture of UO$_{2}$ on U-metal is $\left\langle110\right\rangle$. For a randomly oriented sample the ratio UO$_{2}$(200)/UO$_{2}$(111) $\sim$ 1/3, but to determine the exact texture we should also measure the UO$_{2}$(220), which we have not done in our experiments. No doubt the different textures relate to the fact that the UO$_{2}$ is deposited (either in the sputtering chamber or by air-growth) on mostly epitaxial flat uranium metal. The oxide overlayers we have produced are textured, probably favoring $\left\langle111\right\rangle$ rather than $\left\langle110\right\rangle$. 

Moreover, whereas in the work by Jones et al. \cite{Jones2008}, the air-grown oxide crystallites were found to be about 12 nm in extension when grown on polycrystalline uranium, Figure 8 shows that they are $\sim$ 20 nm in the as-deposited sample and as large as 80 nm (in the vertical direction) in the air-grown sample. 

In addition to the UO$_{2}$(111) and UO$_{2}$(200) described above we observe a reflection at 2$\theta$ = $\sim$32$^{\circ}$ which we assign to a higher UO$_{2+x}$ oxide structure. Higher oxides of UO$_{2}$ are, of course, well known and studied; a review is given by McEachern and Taylor \cite{McEachern1998}, and more specific information on the diffraction spectra is given by Rousseau et al. \cite{Rousseau2006} – see Figures 5 and 6 in that paper. On the basis of that work we assign the reflection at $\sim$32$^{\circ}$ to U$_{3}$O$_{7}$(002). Other U$_{3}$O$_{7}$ reflections are coincident with those of UO$_{2}$ reflections and thus difficult to observe, and the $\sim$32$^{\circ}$ reflection is observed to disappear with hydrogen dosing over about 3 hours for the as-deposited (P$_{H_{2}}$ 2-4 mbar, 80 C) and over about 3.5 hours for the air-grown (PH$_{2}$ 16 mbar, 80 then 140 C) as would be expected if U$_{3}$O$_{7}$ is reduced to UO$_{2.0}$. It should be noted that the reduction of U$_{3}$O$_{7}$(002) occurs over the same period as uranium is consumed indicating that the presence of U$_{3}$O$_{7}$(002) does not appear to impact on the hydriding reaction.

The intensity of the UO$_{2}$(111) reflection is particularly stable for the as-deposited sample (Figure 3a) with no change throughout the hydrogen exposure experiment. For the air-grown sample the picture is more complex with a slight reduction (Figure 5b) in UO$_{2}$ intensity with hydrogen exposure. This might be explained in part by the reduction of any U$_{3}$O$_{7}$ components of the film although one might expect it to be replaced by UO$_{2.0}$ intensity; but in any case this points to considerable changes ocurring within the air-grown oxide film over this period. 

Consideration of the FWHM of the oxide reflections shows a similar picture with the as-deposited (Figure 8a) showing no changes throughout the dosing experiment and with the air-grown oxide showing considerable change (Figure 8b). The FWHM of the air-grown oxide reflections (see Figure 8b) suggest that the crystallite size of UO$_{2}$(111) and (200) decrease steadily (by around 50\%) as a function of hydrogen dose; this happens over the same period where the UO$_{2}$(111) intensity is reduced (Figure 5b)) and is independent of the consumption of uranium. The reason for this loss of intensity is not understood but points to some significant changes within the oxide film during this period. All these changes alter the overlayer, breaking it up and forming a structure with smaller crystallites than in the original sample. At the same time a great acceleration is observed in the rate of metal consumption, shown in Figure 9. In future studies some of these films will be examined with microscopic methods – such as TEM – to try to determine the exact nature of the air-grown oxide and how it reacts to temperatures above 100 C.

Finally, analysis of the position of the UO$_{2}$(111) reflection shows that this d-spacing decreases during hydrogen exposure. The contraction of these d-spacings as a consequence of the hydrogen exposure are significant in size at 0.10\% (as-deposited, 0-1.2 min.bar) and 0.38\% (air-grown, 0-8 min.bar) and would appear to be independent of whether, or how much uranium (underlying the oxide) has been consumed. %While the change in d-spacing for the as-deposited would appear effectively linear, the equivalent rate of change of the equivalent d-spacing in the air-grown sample appears to have a parabolic form (particularly over the long 140 C dose 30-150 min.bar and 200 C dose 150-180 min.bar) with the data suggesting that a contant value will be reached if held at long periods at constant temperature. [Actually, I don’t see this follows from Figure 5d, and suggest we simply delete this sentence – the next one is OK] 
Higher temperatures are responsible for greater changes in the d-spacing.

The most likely explanation for the contraction of the d-spacings is the absorption of hydrogen into the UO$_{2}$ lattice; the change is of an order observed when additional oxygen enters the UO$_{2}$ lattice to form UO$_{2+x}$ \cite{Colmenares1984}. This thesis is consistent with the decreases in d-spacing and with the temperature dependence of the effect. %An alternative explanation for this observation could include a response of the oxide to a changing stress state as the hydrogen consumes the uranium. % or ii) trace oxidant impurities in the gas mixture oxidising the UO$_{2}$ film to UO$_{2}$+x. 
%However, this explanation is discounted since there are no indications of a release of stress later in the experiment indeed for air-grown sample there are indications that the change in d-spacing and the consumption of uranium are not correlated: in this case the uranium is 95\% consumed at a dose of 2.2 min.bar whereas the d-spacing continues to contract out to 8 min.bar. %The second alternative explanation is discounted since the UO$_{2}$(111) intensity is not increased over the experiments as would be expected if the uranium is being corroded by oxidants rather than hydrogen. Indeed, as has been noted above, UO$_{2+x}$ signatures are removed during the dosing period indicating a reducing rather than an oxidising environment.

This last observation, of a decreasing UO$_{2}$(111) d-spacing as a result of H$_{2}$ dosing, is of significance as it provides some evidence at last that hydrogen may enter the UO$_{2}$ lattice and therefore presumably may traverse it during hydrogen corrosion reactions \cite{Glascott2013}. We have observed the effect (and the parabolic-like form) with other samples not reported here, and the observation leaves the way open to use the measurement to probe the kinetics of hydrogen introduction into these UO$_{2}$ films. However, despite providing evidence that the hydrogen is introduced into the UO$_{2}$ lattice, the cumulative evidence here is that the uranium is almost quantitatively consumed (95\% complete) by 0.96 (as-deposited) and 2.2 (air-grown) min.bar early in the reaction and the amount of uranium remaning has no effect on the changing UO$_{2}$(111) d-spacing. This would suggest that the consumption of uranium is unrelated to the absorption of hydrogen into the oxide overlayer. Given that the oxide overlayer is polycrystalline with grains of the order of 20 nm (as-deposited) and 20-80 nm (air-grown), then we can imagine UO$_{2}$ grain boundaries as a viable alternative route for hydrogen to enter the metal. Lastly, it is noteworthy that at high doses on the air-grown sample the UO$_{2}$(200) d-spacing is observed to increase as the UO$_{2}$(111) decreases and this may provide an indication as to the environment(s) in which the hydrogen resides.

\subsection{Observations of UH$_{3}$}

As discussed above, there is overwhelming evidence for consumption of metallic uranium during the hydrogen exposure period; a period where the UO$_{2}$ XRD reflections do not increase in intensity and indeed the U$_{3}$O$_{7}$(002) reflection is removed indicating a reducing environment. However, despite the fact that uranium is consumed the inferred product (either $\alpha$-UH$_{3}$ or $\beta$-UH$_{3}$) has not been observed in the diffraction patterns (Figures 2 and 4). In related work of hydriding exposure of polycrystalline uranium at 80 C (R.M. Harker unpublished) $\beta$-UH$_{3}$(210) was readily identified at a d-spacing of 2.99\AA ($\sim$ 29.8$^{\circ}$ = 2$\theta$, slightly expanded from the tabulated value expected at 30.24$^{\circ}$ = 2$\theta$). In order to be unobservable in the diffraction pattern the compound product would need to be either amorphous or as nanocrystalline as to be effectively amorphous over the XRD length scale.

Transmission Electron Microscopy (Figure 6) of a lift-out foil from a sample exposed to a 10 min.bar hydrogen dose confirms that a compound (presumably UH3) distinct from the oxide is formed in this thin film system, and also confirms that this product is amorphous. 

This supports the view that precipitated UH$_{3}$ in these studies may be unobservable on account of its level of crystallinity. Nanocrystalline $\beta$-UH$_{3}$ was recently observed in a hydrided U-Mo$_{0.18}$ alloy where the product was non-crystalline by XRD but grain sizes of 1-2 nm were determined using Partial Distribution Function analysis \cite{RH12, Havela2016}. UH$_{x}$ has also been observed when H$_{2}$O reacts with uranium metal \cite{Martin2016} showing that various stoichiometries can exist at the interface between metal and oxide.

\section{Conclusions}
Grains of $\alpha$-U(110) are consumed at a faster rate than the minor component grains of $\alpha$-U(002). The rate of consumption of $\alpha$-U(110) in as-deposited oxide at 80 C follows a logarithmic law (Figure 9b)) indicating the generation of an adherent UH$_{3}$ ‘barrier’ that slows subsequent hydriding. The consumption of $\alpha$-U(110) in air-grown oxide is both slower and approximately linear at 80 C. However, when the temperature is increased with the air-grown oxide layer to above 100 C there is a very rapid consumption of uranium (Figure 9).
The available evidence (from changes in the FWHM/inferred crystallite size) suggests that the $\alpha$-U(110) grains are not thinned appreciably in the vertical direction during uranium consumption (Figure 8b)) suggesting a face perpendicular to U(110) is rapidly consumed in these samples as compared to the U(110) face. This leads to a model of consumption in this system of grain-boundary corrosion with the corrosion interface moving laterally into the grains. A consequence of the consumption of uranium is a proportional increase in the d-spacing of U(110), and to a lesser extent the U(002), in the direction of film thickness.

There are considerable differences in the as-deposited and air-grown oxides in these samples and this is responsible for profound differences in how the samples react with hydrogen. Minor reflections of U$_{3}$O$_{7}$(002) are removed during the hydrogen exposure at temperature but this does not appear to impact on the uranium corrosion rates. Also there appears to be considerable changes in the air-grown oxide overlayer when heated to 140 C as indicated by changes in the intensity and FWHM of UO$_{2}$(111) and UO$_{2}$(200). Finally, there is considerable evidence for a contraction of the UO$_{2}$(111) d-spacings during H$_{2}$ exposure providing evidence of hydrogen absorption into the UO$_{2}$ lattice; the contractions are parabolic-like with higher temperatures resulting in faster rates and greater extent of change. This provides evidence that hydrogen may enter the UO$_{2}$ lattice and may thereby traverse the oxide in this way. However this mechanism of hydrogen transport across the oxide appears to not be dominant in this system as the changes in oxide d-spacings change over a longer timescale than is seen for uranium consumption.

Interrogation of these thin film samples by synchrotron X-radiation provides a powerful tool to understand the changes ocurring in the metal and oxide components of this model system.

\section{Acknowledgments} 
AWE is thanked for the supply of low carbon uranium metal material for use as magnetron sputtering target. The authors thank the EPSRC grant code EP/K040375/1 ‘South of England Analytical Electron Microscope’. This work was undertaken in part due to the funding of the DISTINCTIVE consortium EP/L014041/1. We thank Patrick Colomp and the Radiation Protection Group at the ESRF for cooperation during our experiments, Tom Scott of Bristol for his support and encouragement. Additional thanks for Peter Morrall and James Petherbridge for comments on the manuscript.

\section{Appendix}

\noindent    
    \begin{tabularx}{\columnwidth}{|X| X| X| X|}

%\hline
%Test & Test &  Test & Test  \\
%Test & Test &  Test & Test  \\
\hline
 & \multicolumn{2}{|X|}{Thickness(nm)} & \\
Sample	& Metal  &	Oxide & Figures \\
\hline \hline
As-deposited	& 70(5)	& 20(5) &	2, 3, 8, 9 \\ \hline
Air-grown &	$\sim$ 80 &	$\sim$ 80& 4, 5, 8, 9 \\ \hline
STEM &	$\sim$ 45 &	$\leq$ 1 &	6 \\ \hline
EELS &	30(5) &	30(5) &	7 \\ \hline

\end{tabularx}

\section{References}

\bibliography{ReferencesGHL}{}
\bibliographystyle{unsrt}

\end{multicols}
\end{document}